\documentclass[12pt]{article}
\usepackage{ctex}
\usepackage{amssymb,amsmath,color,mathdots}
\usepackage[colorlinks,linkcolor=blue]{hyperref}
\usepackage{geometry}
\usepackage{tabularx}
\usepackage{bm}
\usepackage{authblk}
\usepackage[square, comma, sort&compress, numbers]{natbib}

	\title{\bf Two new classes of projective two-weight linear codes}	
		\author{\small Canze Zhu}
	\author{\small Qunying Liao
		\thanks{Corresponding author.\\
			{E-mail. qunyingliao@sicnu.edu.cn (Q. Liao), ~canzezhu@163.com (C. Zhu).}	\\				
			{~Supported by National Natural Science Foundation of China (Grant No. 12071321).}}
		}
	\affil[]{\small (College of Mathematical Science, Sichuan Normal University, Chengdu Sichuan, 610066)}
	\date{}
	\newtheorem{theorem}{Theorem}[section]
	
	\newtheorem{lemma}{Lemma}[section]
	\newtheorem{example}{Example}[section]
	
	\newtheorem{corollary}{Corollary}[section]
	\newtheorem{remark}{Remark}[section]

	\catcode`@=11 \@addtoreset{equation}{section} \catcode`@=12
\begin{document}
	\maketitle
	{\bf Abstract.}
	{\small
	 In this paper, for an odd prime $p$, several classes of two-weight linear codes over the finite field $\mathbb{F}_p$ are constructed from the defining sets, and then their complete weight distributions are determined by employing character sums. These codes can be suitable for applications in secret sharing schemes. Furthermore, two new classes of projective two-weight codes are obtained, and then two new classes of strongly regular graphs are given. }\\
	

	\section{Introduction}
	Let $\mathbb{F}_{p^m}$ be the finite field with $p^m$ elements and $\mathbb{F}_{p^m}^*=\mathbb{F}_{p^m}\backslash \{0\}$, where $p$ is an odd prime and $m$ is a positive integer. An $[n,k,d]$ linear code $\mathcal{C}$ over $\mathbb{F}_p$ is a $k$-dimensional subspace of $\mathbb{F}_p^n$ with minimum (Hamming) distance $d$ and length $n$. 
	The dual code of $\mathcal{C}$ is defined as 
	$$	 \mathcal{C}^{\perp}=\{\mathbf{c}^{\perp}\in\mathbb{F}_p^n~|~\langle\mathbf{c}^{\perp},\mathbf{c}\rangle=0~ \text{for any}~\mathbf{c}\in\mathcal{C}\}.$$
	Clearly, the dimension of $\mathcal{C}^{\perp}$ is $n-k$. A linear code $\mathcal{C}$ is said to be projective if the minimum distance of $\mathcal{C}^{\perp}$  is greater than or equal to $3$.

	Let $A_i\ (i=1,2,\ldots,n)$ be the number of codewords with Hamming weight $i$ in $\mathcal{C}$, then the weight distribution  of $\mathcal{C}$ is defined by the polynomial
		$$1+A_1z+A_2z^2+\cdots+A_nz^n. $$ $\mathcal{C}$ is called a $t$-weight code if the number of nonzero $A_i$ in the sequence $(A_1,A_2,\ldots,A_n)$ is equal to $t$. In addition, the complete weight enumerator of a codeword $\mathbf{c}$ is the monomial
		$$w(\mathbf{c})=w_0^{t_0}w_1^{t_1}\cdots w_{p-1}^{t_{p-1}}$$
	in the variables $w_0,w_1,\ldots,w_{p-1}$, where $t_i\ (0\le i\le p-1)$ denotes the number of components of $\mathbf{c}$ equal to $i$. The complete weight enumerator of $\mathcal{C}$ is defined to be
	\begin{center}
		$W(\mathcal{C})=\sum\limits_{\mathbf{c}\in\mathcal{C}}w(\mathbf{c})$.
	\end{center} 
	
	 The complete weight enumerator is an important parameter for a linear code, obviously, the weight distribution  can be deduced from the complete weight enumerator. In addition, the weight distribution  for $\mathcal{C}$ can be applied to determine the  capability for both error-detection and error-correction \cite{KT2007}.  
	 
	Linear codes over finite fields are applied in data storage devices, computer and communication systems, and so on. Since few-weight linear codes have been better applications in secret sharing schemes \cite{JY2006,CC2005},  association schemes \cite{AC1984}, authentication codes \cite{CD2005}, and so on.  A number of two-weight or three-weight linear codes have been constructed \cite{KD2014, KD2015, SY2015, ZH2015, ZH2016, ZH20161, ZH20162, GJ2019, CL2016, GL2018, CT2016, ZZ2015, Z, CS2019,ZH2017,ZH2021,CD2007,CD2008,CD2015}. In particular, projective two-weight codes are very precious as they are closely related to finite projective spaces, strongly regular graphs and combinatorial
	designs \cite{RC1986,CD2018,P1972}. However, projective two-weight codes are rare and only a few classes are known \cite{RC1986,ZH20161,ZH2017,ZH2021,CD2007,CD2008,CD2015}.
	 
        In 2007, Ding, et al. gave a construction for linear codes via the trace function from a defining set \cite{CD2007}. Let 
        $D=\{ d_1, d_2,\ldots, d_n \}\subseteq\mathbb{F}_{p^{m}}^*$ and $\mathrm{Tr}$ denote the trace function from $\mathbb{F}_p^{m}$ onto $\mathbb{F}_p$, a $p$-ary linear code is defined by
	  \begin{align*}
	  \mathcal{C}_D=\big\{\mathbf{c}(x)=\big(\mathrm{Tr}(xd_1),\mathrm{Tr}(xd_2),\ldots,\mathrm{Tr}(xd_n)\big)~\big{|}~x\in\mathbb{F}_p^{m}\big\}.
	  \end{align*} 
	  Motivated by the above construction, Li, et al. defined a linear code by
	  	 	\begin{align}\label{CD}
	  	 	\mathcal{C}_{D}=\big\{\mathbb{c}(a,b)=\big(\mathrm{Tr}(ax+by)\big)_{(x,y)\in D}~\big{|}~ (a,b)\in\mathbb{F}_{p^m}\times\mathbb{F}_{p^m}\big\}
	  	 	\end{align}
	 with $D\subseteq \mathbb{F}_{p^m}^2$ \cite{CL2016} . 

	In this paper, for a given positive integer $d$ and $\tilde{D}\subseteq \mathbb{F}_{p^m}^2$, the linear code $\mathcal{C}_{\tilde{D}}$ is defined by 
	\begin{align}\label{C_D}
	\mathcal{C}_{\tilde{D}}=\big\{\big(\mathrm{Tr}(ayx^d+bx)\big)_{(x,y)\in {\tilde{D}}}~\big{|}~(a,b)\in\mathbb{F}_{p^m}\times\mathbb{F}_{p^m}\big\}.
	\end{align}	
	  At the same time, we also consider the defining sets
	  \begin{align}\label{D_2}
	 D^*
	 =&\big\{(x,y)\in\mathbb{F}_{p^m}^{*}\times\mathbb{F}_{p^{m}}^*~\big{|}~ \mathrm{Tr}(yx^{d+1})=0\big\}
	 \end{align} and 
	 \begin{align}\label{D_1}
	 D_\lambda
	 =&\big\{(x,y)\in\mathbb{F}_{p^m}^{*}\times\mathbb{F}_{p^{m}}~\big{|}~ \mathrm{Tr}(yx^{d+1})=\lambda\big\},
	 \end{align}
	where $\lambda\in\mathbb{F}_p$. For $\tilde{D}$ = $D^*$ or $D_\lambda$,
	the parameters and the complete weight enumerator of $\mathcal{C}_{\tilde{D}}$ will be determined based on character sums. In particular, $\mathcal{C}_{\tilde{D}}$ is suitable for applications in secret sharing schemes. Furthermore, the punctured code for $\mathcal{C}_{D_0}$ or $\mathcal{C}_{D^*}$ is a projective two-weight code, by which a strongly regular graph with new parameters can be derived.
	
	The paper is organized as follows. In section 2, some related basic notations and results for character sums are given. In section 3, we present the parameters for several classes of two-weight linear codes, and then obtain two classes of projective two-weight codes.  In section 4, the proofs for main results are given. In section 5, we show that these codes can be applicated for secret sharing schemes, and some strongly regular graphs with new parameters are derived basing on projective two-weight codes. In section 6, we conclude the whole paper.
	\section{Preliminaries}
    \indent An additive character $\chi$ of $\mathbb{F}_{p^m}$ is a function from $\mathbb{F}_{p^m}$ to the multiplicative group $U=\{u\ |\ |u|=1,\ u\in\mathbb{C}\}$, such that $\chi(x+y)=\chi(x)\chi(y)$ for any $x,\ y\in\ \mathbb{F}_{p^m}$. For each $b\in \mathbb{F}_{p^m}$, the function
    \begin{center}
    	$\chi_b(x)=\zeta_p^{\mathrm{Tr}(bx)}$~~$(x\in \mathbb{F}_{p^m})$
    \end{center}
    is an additive character of $\mathbb{F}_{p^m}$. When $b=0$, i.e, $\chi_0(c)=1$ for any $c\in \mathbb{F}_{p^m}$, $\chi_0$ is called the trivial additive character of $\mathbb{F}_{p^m}$. The character $\chi:=\chi_1$ is called the canonical additive character of $\mathbb{F}_{p^m}$ and every
    additive character of $\mathbb{F}_{p^m}$ can be written as $\chi_b(x)=\chi(bx)$. The orthogonal property for additive characters is given by
    \begin{align*}
    	\sum_{x\in \mathbb{F}_{p^m}}\zeta_p^{\mathrm{Tr}(bx)}=\begin{cases}
    		p^m,\quad& ~b=0;\\
    		0,\quad &\text{otherwise}.
    	\end{cases}
    \end{align*}     
    We extend the quadratic characters $\eta_m$ of $\mathbb{F}_{p^m}^*$ by letting $\eta_m(0)=0$, then the quadratic Gauss sums $G_m$ over $\mathbb{F}_{p^m}$ is defined to be
    \begin{center}
    	$G_m=\sum\limits_{c\in \mathbb{F}_{p^m}^*}\eta_m(c)\chi(c)=\sum\limits_{c\in \mathbb{F}_{p^m}}\eta_m(c)\chi(c)$.
    \end{center}
    
      Now, some properties for quadratic characters and quadratic Gauss sums are given as follows.    	
    \begin{lemma}[\cite{KD2015}, Lemma 7]
    	For $x\in\mathbb{F}_p^*$,  
    	\begin{align*}
    		\eta_m(x)=\begin{cases}
    			1,\quad& 2|m;\\
    			\eta_1(x),\quad &\text{otherwise}.
    		\end{cases}
    	\end{align*}     	
    \end{lemma} 
    \begin{lemma}[\cite{RL97}, Theorem 5.15]\label{GS}
    	For the Gauss sums $G_m$ over $\mathbb{F}_{p^m}$,
    	\begin{align*}
    		G_m=(-1)^{m-1}(\sqrt{-1})^\frac{(p-1)^2m}{4}p^{\frac{m}{2}}.
    	\end{align*}
    \end{lemma}
    \begin{lemma}[\cite{RL97}, Theorem 5.33]\label{L1.2}
    	For $f(x)=a_2x^2+a_1x+a_0\in \mathbb{F}_q[x]$ with $a_2\neq0$, 
    	\begin{align*}
    	\sum_{c\in \mathbb{F}_q}\chi(f(c))=G_m\eta_m(a_2)\chi(a_0-a_1^2(4a_2)^{-1}).
    	\end{align*}
    \end{lemma}

The Pless power moments are useful for determining the minimum distance of the dual for a
linear code.


\begin{lemma}[\cite{WV}, p.259, The Pless power moments]\label{l12}
	For an $[n, k, d]$ code $\mathcal{C}$ over $\mathbb{F}_p$ with the weight distribution $(1, A_1,\ldots, A_n)$, suppose that
	the weight distribution  of  its dual code is $(1, A_1^{\bot},\ldots,A_n^{\bot})$, then the first two Pless power
	moments are
	\begin{align*}
	\sum_{j=0}^{n}A_j=p^{k}
	\end{align*}and
	\begin{align*}
	\sum_{j=0}^{n}jA_j=p^{k-1}(pn-n- A_1^{\bot}),
	\end{align*}	
	respectively. Furthermore, if $(0,0)\notin {D_\lambda}$, then $A_1^{\bot}=0$. 	
\end{lemma}
    \section{Our Main Results}
	\begin{theorem}\label{t1}
		For any integer $m\ge2$, suppose that $D_0$ and $\mathcal{C}_{D_0}$ are given by $(\ref{D_1})$ and $(\ref{C_D})$, respectively, then $\mathcal{C}_{D_0}$ is a $\big{[}p^{2m-1}-p^{m-1}, 2m,(p-1)(p^{m-1}-1)p^{m-1}\big{]}$ code with the weight distribution  in Table $1$, and the complete weight enumerator is 
		\begin{align}\label{w1}
		\begin{aligned}
		\mathrm{W}(\mathcal{C}_{D_0})=& w_0^{p^{2m-1}-p^{m-1}}+(p^m-p^{m-1}+1)(p^m-1)w_0^{p^{2m-2}-p^{m-1}}\prod_{i\in \mathbb{F}_p^*}w_i^{p^{2m-2}}\\
		&+p^{m-1}(p^{m}-1)w_0^{p^{2m-2}+(p-2)p^{m-1}}\prod_{i\in \mathbb{F}_p^*}w_i^{p^{2m-2}-p^{m-1}}.
		\end{aligned}
		\end{align}	

	\begin{center} Table $1$. the weight distribution  of $\mathcal{C}_{D_0}$
				
		\begin{tabular}{|p{5cm}<{\centering}| p{7cm}<{\centering}|}
			\hline   weight $w$	                     &   frequency $A_w$                    \\ 
			\hline       $0$	                     &  $1$                                    \\ 
			\hline   $(p-1)p^{2m-2}$   &  $(p^m-1)(p^m-p^{m-1}+1)$        \\ 
			\hline   $(p-1)(p^{m-1}-1)p^{m-1}$	                     &      $(p^{m}-1) p^{m-1}$  \\         
			\hline
		\end{tabular} 
	\end{center} 
	\end{theorem}
	
	\begin{theorem}\label{t2}
		For any integer $m\ge2$, suppose that $D^*$ and $\mathcal{C}_{D^*}$ are given by $(\ref{D_2})$ and $(\ref{C_D})$, respectively, then $\mathcal{C}_{D^*}$ is a $\big{[}p^{2m-1}-p^{m}-p^{m-1}+1, 2m,(p-1)(p^{m-1}-2)p^{m-1}\big{]}$ code with the weight distribution  in Table $2$, and the complete weight enumerator is 
		\begin{align}\label{w2}
		\begin{aligned}
		\mathrm{W}(\mathcal{C}_{D^*})=&w_0^{p^{2m-1}-p^{m}-p^{m-1}+1}+(p^m-p^{m-1}+2)(p^m-1)w_0^{p^{2m-2}-2p^{m-1}+1}\prod_{i\in \mathbb{F}_p^*}w_i^{(p^{m-1}-1)p^{m-1}}\\
		&+(p^{m-1}-1)(p^{m}-1)w_0^{p^{2m-2}+(p-3)p^{m-1}+1}\prod_{i\in \mathbb{F}_p^*}w_i^{(p^{m-1}-2)p^{m-1}}.
		\end{aligned}
		\end{align}	
		
		\begin{center} Table $2$. the weight distribution  of $\mathcal{C}_{D^*}$
			
			\begin{tabular}{|p{5cm}<{\centering}| p{7cm}<{\centering}|}
				\hline   weight $w$	                     &   frequency $A_w$                    \\ 
				\hline       $0$	                     &  $1$                                    \\ 
				\hline   $(p-1)(p^{m-1}-1)p^{m-1}$   &  $(p^m-1)(p^m-p^{m-1}+2)$        \\ 
				\hline   $(p-1)(p^{m-1}-2)p^{m-1}$	                     &      $ (p^{m}-1)(p^{m-1}-1)$  \\         
				\hline
			\end{tabular} 	\\
		\end{center} 
	\end{theorem}
 
	

	\begin{theorem}\label{t3}
		For any integer $m\ge2$ and $\lambda\in \mathbb{F}_p^*$, suppose that $D_\lambda$ and $\mathcal{C}_{D_\lambda}$ are given by  $(\ref{D_1})$ and $(\ref{C_D})$, respectively, then $\mathcal{C}_{D_\lambda}$ is a $\big[p^{2m-1}-p^{m-1}, 2m,(p^{m}-p^{m-1}-2)p^{m-1}\big]$ code with the weight distribution  in Table $3$, and the complete weight enumerator is 
		\begin{align}\label{w3}
		\begin{aligned}
		\mathrm{W}(\mathcal{C}_{D_\lambda})=& w_0^{p^{2m-1}-p^{m-1}}+(p^{m-1}+1)(p^m-1)w_0^{p^{2m-2}-p^{m-1}}\prod_{i\in \mathbb{F}_p^*}w_i^{p^{2m-2}}\\
		&+\frac{p-1}{2}p^{m-1}(p^{m}-1)\sum_{\substack{j\in\mathbb{F}_p^*\\\eta_1(-\lambda j)=-1}}
		\prod_{i\in\mathbb{F}_p}w_i^{p^{2m-2}+\eta_1(i^2-4\lambda j)p^{m-1}}\\
		&+\frac{p-1}{2}p^{m-1}(p^{m}-1)\sum_{\substack{j\in\mathbb{F}_p^*\\\eta_1(-\lambda j)=1}}\prod_{i\in \mathbb{F}_p}w_i^{p^{2m-2}+\eta_1(i^2-4\lambda j)p^{m-1}}.		
		\end{aligned}
		\end{align}	
		
		\begin{center} Table $3$. the weight distribution  of $\mathcal{C}_{D_\lambda}$
			
			\begin{tabular}{|p{5cm}<{\centering}| p{7cm}<{\centering}|}
				\hline   weight $w$	                     &   frequency $A_w$                    \\ 
				\hline       $0$	                     &  $1$                                    \\ 
				\hline   $(p-1)p^{2m-2}$   &  $(\frac{p+1}{2}p^{m-1}+1)(p^m-1)$        \\ 
				\hline   $(p^{m}-p^{m-1}-2)p^{m-1}$	                     &      $ \frac{p-1}{2}p^{m-1}(p^{m}-1)$  \\         
				\hline
			\end{tabular} \\
		\end{center} 
	\end{theorem}

 In the following, we give the  punctured codes for $\mathcal{C}_{D_0}$, $\mathcal{C}_{D^*}$ and $\mathcal{C}_{D_{\lambda}}$, respectively, and then obtain two classes of projective two-weight codes.
 
For any  $c\in \mathbb{F}_{p}^{*}$, if $(p-1)\mid d$, then
\begin{align}\label{ss1}
\mathrm{Tr}((cy)(cx)^{d+1})=c^{d+2}\mathrm{Tr}(yx^{d+1})=c^{2}\mathrm{Tr}(yx^{d+1})
\end{align} 
and 
\begin{align}\label{ss2}
\mathrm{Tr}\big(a(cy)(cx)^{d}+b(cx)\big)=c\mathrm{Tr}(ayx^d+bx)\quad ~(a,b\in\mathbb{F}_{p^m}).
\end{align}
From $(\ref{ss1})$-$(\ref{ss2})$, $D_0$ and $D^*$ can be expressed as
\begin{align}\label{pd1}
{D_0}=\cup_{c\in\mathbb{F}_p^{*}}c\overline{D_0}
\end{align}
and
\begin{align}\label{pd2}
{D^*}=\cup_{c\in\mathbb{F}_p^{*}}c\overline{D^*},
\end{align}
respectively, where $\overline{D_0}\subsetneq{D_0}$ and $\overline{D^*}\subsetneq{D^*}$, and then both $\mathcal{C}_{\overline{D_0}}$ and $\mathcal{C}_{\overline{D^*}}$ defined by $(\ref{C_D})$ are just the punctured versions of $\mathcal{C}_{D_0}$ and $\mathcal{C}_{D^*}$, respectively, whose parameters are given in the following corollaries.

\begin{corollary}\label{c1}
	For any integer $m\ge2$, suppose that $\overline{D_0}$ and $\mathcal{C}_{\overline{D_0}}$ are given by $(\ref{pd1})$ and $(\ref{C_D})$, respectively, then $\mathcal{C}_{\overline{D_0}}$ is a $\big{[}\frac{p^{2m-1}-p^{m-1}}{p-1}, 2m,(p^{m-1}-1)p^{m-1}\big{]}$ code with the weight distribution  in Table $4$. 
	
	\begin{center} Table $4$. the weight distribution  of $\mathcal{C}_{\overline{D_0}}$
		
		\begin{tabular}{|p{5cm}<{\centering}| p{7cm}<{\centering}|}
			\hline   weight $w$	                     &   frequency $A_w$                    \\ 
			\hline       $0$	                     &  $1$                                    \\ 
			\hline   $p^{2m-2}$   &  $(p^m-p^{m-1}+1)(p^m-1)$        \\ 
			\hline   $(p^{m-1}-1)p^{m-1}$	                     &      $ p^{m-1}(p^{m}-1)$  \\         
			\hline
		\end{tabular} 
	\end{center} 
\end{corollary}

\begin{corollary}\label{c2}
	For any integer $m\ge2$, suppose that  $\overline{D^*}$ and $\mathcal{C}_{\overline{D^*}}$ are given by $(\ref{C_D})$ and $(\ref{pd2})$, respectively, then $\mathcal{C}_{\overline{D^*}}$ is a $\big{[}\frac{p^{2m-1}-p^{m}-p^{m-1}+1}{p-1}, 2m,(p^{m-1}-2)p^{m-1}\big{]}$ code with the weight distribution  in Table $5$.
	
	\begin{center} Table $5$. the weight distribution  of $\mathcal{C}_{\overline{D^*}}$
		
		\begin{tabular}{|p{5cm}<{\centering}| p{7cm}<{\centering}|}
			\hline   weight $w$	                     &   frequency $A_w$                    \\ 
			\hline       $0$	                     &  $1$                                    \\ 
			\hline   $(p^{m-1}-1)p^{m-1}$   &  $(p^m-p^{m-1}+2)(p^m-1)$        \\ 
			\hline   $(p^{m-1}-2)p^{m-1}$	                     &      $ (p^{m}-1)(p^{m-1}-1)$  \\         
			\hline
		\end{tabular} 
	\end{center} 
\end{corollary}

By Lemma $\ref{l12}$, we can get that the minimum diatance for $\mathcal{C}_{\overline{D_0}}^{\perp}$ or $\mathcal{C}_{\overline{D^*}}^{\perp}$ is $3$, thus we have the following corollaries.
\begin{corollary}\label{c4}
	For any integer $m\ge2$, suppose that $\overline{D_0}$ and $\mathcal{C}_{\overline{D_0}}$ are given by $(\ref{pd1})$ and $(\ref{C_D})$, respectively, then $\mathcal{C}_{\overline{D_0}}^{\perp}$ is a $\big{[}\frac{p^{2m-1}-p^{m-1}}{p-1},\frac{p^{2m-1}-p^{m-1}}{p-1}-2m,3\big{]}$ code, and $\mathcal{C}_{D_0}$ is a projective two-weight code.
\end{corollary}

\begin{corollary}\label{c5}
	For any integer $m\ge2$, suppose that $\overline{D^*}$ and $\mathcal{C}_{\overline{D^*}}$ are given by $(\ref{pd2})$ and $(\ref{C_D})$, respectively, then $\mathcal{C}_{\overline{D^{*}}}^{\perp}$ is a $\big{[}\frac{p^{2m-1}-p^m-p^{m-1}-1}{p-1},\frac{p^{2m-1}-p^m-p^{m-1}-1}{p-1}-2m,3\big{]}$ code, and $\mathcal{C}_{D^*}$ is a projective two-weight code.
\end{corollary}

    For $\alpha=-1$ and any even $d$, we have
	\begin{align*}
	\mathrm{Tr}((\alpha y)(\alpha x)^{d+1})=\mathrm{Tr}(yx^{d+1})
	\end{align*} 
	and 
	\begin{align*}
	\mathrm{Tr}\big(a(\alpha y)(\alpha x)^{d}+b(\alpha x)\big)=\alpha\mathrm{Tr}(ayx^d+bx)\quad (a,b\in\mathbb{F}_{p^m}).
	\end{align*}
   Hence, ${D_\lambda}$ can be expressed as
\begin{align}\label{pd3}
{D_\lambda}=\overline{D_\lambda}\cup\alpha\overline{D_\lambda},
\end{align}
where $\overline{D_\lambda}\subsetneq{D_\lambda}$. Then $\mathcal{C}_{\overline{D_\lambda}}$ defined by $(\ref{C_D})$ is just the punctured version  of $\mathcal{C}_{D_\lambda}$, and the parameters for $\mathcal{C}_{\overline{D_\lambda}}$ are given in the following
	\begin{corollary}\label{c3}
	For any integer $m\ge2$ and $\lambda\in \mathbb{F}_p^*$, suppose that $\overline{D_\lambda}$ and $\mathcal{C}_{\overline{D_\lambda}}$ are given by $(\ref{pd3})$ and $(\ref{C_D})$, respectively, then $\mathcal{C}_{\overline{D_\lambda}}$ is a $\big[\frac{p^{2m-1}-p^{m-1}}{2}, 2m,\frac{(p^{m}-p^{m-1}-2)p^{m-1}}{2}\big]$ code with the weight distribution  in Table $6$.
	
	\begin{center} Table $6$. the weight distribution  of $\mathcal{C}_{D_\lambda}$
		
		\begin{tabular}{|p{5cm}<{\centering}| p{7cm}<{\centering}|}
			\hline   weight $w$	                     &   frequency $A_w$                    \\ 
			\hline       $0$	                     &  $1$                                    \\ 
			\hline   $\frac{p-1}{2}p^{2m-2}$   &  $(\frac{p+1}{2}p^{m-1}+1)(p^m-1)$        \\ 
			\hline   $\frac{(p^{m}-p^{m-1}-2)p^{m-1}}{2}$	                     &      $ \frac{p-1}{2}p^{m-1}(p^{m}-1)$  \\         
			\hline
		\end{tabular} \\
	\end{center} 
	\end{corollary}

	\begin{example}\label{e1}
For $p=3$, $m=2$, $d=2$ and $\lambda=1$, by $(\ref{C_D})$-$(\ref{D_1})$, using MAGMA program, we obtain $\mathcal{C}_{{D_0}}=[24,4,12]$ with the weight distribution $1 +24z^{12}+56z^{18}$, $\mathcal{C}_{{D^*}}=[16,4,6]$ with the weight distribution $1 +16z^{6}+64z^{12}$ and $C_{D_\lambda}=[24,4,12]$ with the weight distribution $1 +24z^{12}+56z^{18}$, which are accordant with Theorems $\ref{t1}$-$\ref{t3}$, respectively. 
	\end{example}	
	\begin{example}
	For $p=3$, $m=2$, $d=2$ and $\lambda=1$, by Example $\ref{e1}$, we can get $\mathcal{C}_{\overline{D_0}}=[12,4,6]$ with the weight distribution $1 +24z^{6}+56z^{9}$,  $\mathcal{C}_{\overline{D^*}}=[8,4,3]$ with the weight distribution $1 +16z^{3}+64z^{6}$ and  $\mathcal{C}_{\overline{D_\lambda}}=[12,4,6]$ with the weight distribution $1 +24z^{6}+56z^{9}$. All these codes are almost optimal according to the Griesmer bound \cite{JH1960}.
	\end{example}
    \begin{remark}
	Easily, $\mathcal{C}_{D^*}$ is obtained by deleting some components of codewords in $\mathcal{C}_{D_0}$.
	\end{remark}
			
	\section{The Proofs for Main Results}
	\subsection{Some auxiliary lemmas}
    In this subsection, Lemmas $\ref{Length}$-$\ref{ML*}$ are useful for calculating the length and the weights, and Lemma $\ref{l2}$ is important to calculate the weight distribution.
    	\begin{lemma}\label{Length}
    	For $\lambda\in\mathbb{F}_p$, we have
    \begin{align}\label{NDL1}
    	\#  D_\lambda=p^{2m-1}-p^{m-1}
    \end{align}
    and
    \begin{align}\label{NDL*}
    		\#  D^*=p^{2m-1}-p^m-p^{m-1}+1.
    \end{align}
    \end{lemma}

	{\bf Proof}. By calculating directly, we have
\begin{align*}
\#  D_\lambda
=& \sum_{x\in\mathbb{F}_{p^m}^*}\sum_{y\in\mathbb{F}_{p^m}}
\big(p^{-1}\sum_{z_1\in\mathbb{F}_p}\zeta_p^{z_1(\mathrm{Tr}(x^{d+1}y)-\lambda)}\big)\\
=& p^{-1}\sum_{x\in\mathbb{F}_{p^m}^*}\sum_{y\in\mathbb{F}_{p^m}}
\big(\sum_{z_1\in\mathbb{F}_p^*}\zeta_p^{z_1(\mathrm{Tr}(yx^{d+1})-\lambda)}+1\big)\\
=& p^{m-1}(p^m-1)+p^{-1}\sum_{z_1\in\mathbb{F}_p^*}\zeta_p^{-z_1\lambda}\sum_{x\in\mathbb{F}_{p^m}^*}\sum_{y\in\mathbb{F}_{p^m}}
\zeta_p^{\mathrm{Tr}(z_1x^{d+1}y)}\\
=&p^{2m-1}-p^{m-1}.
\end{align*}
It follows from  $(\ref{D_2})$-$(\ref{D_1})$ that 
\begin{align*}
	\#D^*=\#D_0-(p^m-1)=p^{2m-1}-p^m-p^{m-1}+1.
\end{align*}$\hfill\Box$
	\begin{lemma}\label{ML}
		For any positive integer $d$, suppose that $\lambda,\lambda_1\in \mathbb{F}_{p}$, $(a,b)\in \mathbb{F}_{p^m}\times\mathbb{F}_{p^m}\backslash\{(0,0)\}$, and
		\begin{align*}
		N_{\lambda,\lambda_1}(a,b)=\{(x,y)\in \mathbb{F}_{p^m}^*\times \mathbb{F}_{p^m}~\big{|}~\mathrm{Tr}(x^{d+1}y)=\lambda~\text{and}~\mathrm{Tr}(ax^dy+bx)=\lambda_1\}.
		\end{align*}	
		Then the following assertions hold.
		
		$(1)$ For $\lambda=\lambda_1=0$,
		\begin{align}\label{5.1}
			\begin{aligned}
				&\#	N_{0,\lambda_1}(a,b)\\=&
				\begin{cases}
					p^{2m-2}-p^{m-1},~&\text{if}~a=0~\text{and}~b\neq0,~\text{or}~\mathrm{Tr}(ab)\neq 0;\\
					p^{2m-2}+(p-2)p^{m-1},~&\text{if}~a\neq0~\text{and}~\mathrm{Tr}(ab)=0.
				\end{cases}
			\end{aligned}
		\end{align}
		
	$(2)$ For $\lambda=0$ and $\lambda_1\neq0$, 
		\begin{align}\label{5.2}
		\begin{aligned}
			&\#	N_{0,\lambda_1}(a,b)\\=&
		\begin{cases}
		p^{2m-2},~&\text{if}~a=0~\text{and}~b\neq0,~\text{or}~a\neq0~\text{and}~\mathrm{Tr}(ab)\neq 0;\\
		p^{2m-2}-p^{m-1},~&\text{if}~a\neq0~\text{and}~\mathrm{Tr}(ab)=0.
		\end{cases}
		\end{aligned}
		\end{align}
		
		$(3)$ For $\lambda\neq0$ and $\lambda_1=0$,			
		\begin{align}\label{5.3}
		\begin{aligned}
			&\#	N_{0,\lambda_1}(a,b)\\=&
		\begin{cases}
		p^{2m-2}-p^{m-1},~&\text{if}~a=0~\text{and}~b\neq0,~\text{or}~a\neq0~\text{and}~\mathrm{Tr(ab)=0};\\
		p^{2m-2}+\eta_1(-\lambda\mathrm{Tr}(ab))p^{m-1},~&\text{if}~\mathrm{Tr}(ab)\neq 0.
		\end{cases}
		\end{aligned}
		\end{align}

		$(4)$ For $\lambda\neq0$ and $\lambda_1\neq0$,		
		\begin{align}\label{5.4}
		\begin{aligned}
			&\#	N_{0,\lambda_1}(a,b)\\=&
		\begin{cases}
		p^{2m-2},~&\text{if}~a=0~\text{and}~b\neq0,~\text{or}~a\neq0~\text{and}~\mathrm{Tr(ab)=0};\\
		p^{2m-2}+\eta_1(\lambda_1^2-4\lambda\mathrm{Tr}(ab))p^{m-1},~&\text{if}~a\neq0~\text{and}~\mathrm{Tr(ab)\neq 0}.
		\end{cases}
		\end{aligned}
		\end{align}							
	\end{lemma}

	{\bf Proof}.~ By calculating directly, we have
	\begin{align*}
		&\#	N_{\lambda,\lambda_1}(a,b)\\
		=& \sum_{x\in\mathbb{F}_{p^m}^*}\sum_{y\in\mathbb{F}_{p^m}}
		\bigg(p^{-1}\sum_{z_1\in\mathbb{F}_p}\zeta_p^{z_1(\mathrm{Tr}(x^{d+1}y)-\lambda)}\bigg)
		\bigg(p^{-1}\sum_{z_2\in\mathbb{F}_p}\zeta_p^{z_2(\mathrm{Tr}(ayx^d+bx)-\lambda_1)}\bigg)\\
		=& p^{-2}\sum_{x\in\mathbb{F}_{p^m}^*}\sum_{y\in\mathbb{F}_{p^m}}
		\big(\sum_{z_1\in\mathbb{F}_p^*}\zeta_p^{z_1(\mathrm{Tr}(yx^{d+1})-\lambda)}+1\big)
		\big(\sum_{z_2\in\mathbb{F}_p^*}\zeta_p^{z_2(\mathrm{Tr}(ax^dy+bx)-\lambda_1)}+1\big)\\
		=&p^{m-2}(p^m-1)+p^{-2}\sum_{z_2\in\mathbb{F}_p^*}\zeta_p^{-z_2\lambda_1}\sum_{x\in\mathbb{F}_{p^m}^*}\sum_{y\in\mathbb{F}_{p^m}}
		\zeta_p^{\mathrm{Tr}(z_2(ayx^d+bx))}\\
		&+ p^{-2}\sum_{z_1\in\mathbb{F}_p^*}\sum_{z_2\in\mathbb{F}_p^*}	  \zeta_p^{-(z_1\lambda+z_2\lambda_1)}\sum_{x\in\mathbb{F}_{p^m}^*}\zeta_p^{\mathrm{Tr}(z_2bx)}\sum_{y\in\mathbb{F}_{p^m}}
		\zeta_p^{\mathrm{Tr}((z_1x+z_2a)x^dy)}\\
		=& p^{m-2}(p^m-1)+\Omega_1+\Omega_2.
	\end{align*}

	For $\Omega_1+\Omega_2$, we have the following two cases.

	{\bf Case 1}. For $a=0$ and $b\neq0$, 
	\begin{align*}
		&\Omega_1+\Omega_2\\
		=& p^{-2}\sum_{z_2\in\mathbb{F}_p^*}\zeta_p^{-z_2\lambda_1}\sum_{x\in\mathbb{F}_{p^m}^*}\sum_{y\in\mathbb{F}_{p^m}}\zeta_p^{\mathrm{Tr}(z_2bx)}\\
		&+ p^{-2}\sum_{z_1\in\mathbb{F}_p^*}\sum_{z_2\in\mathbb{F}_p^*}	 \zeta_p^{-(z_1\lambda+z_2\lambda_1)}\sum_{x\in\mathbb{F}_{p^m}^*}\zeta_p^{\mathrm{Tr}(z_2bx)}\sum_{y\in\mathbb{F}_{p^m}}\zeta_p^{\mathrm{Tr}((z_1x^{d+1}y)}\\
		=& -p^{m-2}\sum_{z_2\in\mathbb{F}_p^*}\zeta_p^{-(z_2\lambda_1)}\\
		=&\begin{cases}
			-(p-1)p^{m-2},~&~\lambda_1=0;\\
			p^{m-2},   ~&\text{otherwise}.
		\end{cases}
	\end{align*}
	
	{\bf Case 2}. For $a\neq0$, one has
	\begin{align*}
		\Omega_1=0,
	\end{align*}
	and
	\begin{align*}
		\Omega_2
		=& p^{-2}\sum_{z_1\in\mathbb{F}_p^*}\sum_{z_2\in\mathbb{F}_p^*}	 \zeta_p^{-(z_1\lambda+z_2\lambda_1)}\sum_{x\in\mathbb{F}_{p^m}^*}\zeta_p^{\mathrm{Tr}(z_2bx)}\sum_{y\in\mathbb{F}_{p^m}}
		\zeta_p^{\mathrm{Tr}((z_1x+z_2a)x^dy)}\\
		=& p^{-2}\sum_{z_1\in\mathbb{F}_p^*}\sum_{z_2\in\mathbb{F}_p^*}	 \zeta_p^{-(z_1\lambda+z_2\lambda_1)}\sum_{z_1x+z_2a=0}\zeta_p^{\mathrm{Tr}(z_2bx)}\sum_{y\in\mathbb{F}_{p^m}}
		\zeta_p^{\mathrm{Tr}((z_1x+z_2a)x^dy)}\\	
		=& p^{m-2}\sum_{z_1\in\mathbb{F}_p^*}\sum_{z_2\in\mathbb{F}_p^*}
		\zeta_p^{-(z_1\lambda+z_2\lambda_1)}\zeta_p^{-z_1^{-1}z_2^2\mathrm{Tr(ab)}}.
	\end{align*}		
	If $\mathrm{Tr}(ab)=0$, we can obtain
	\begin{align*}
		\Omega_2&=p^{m-2}\sum_{z_1\in\mathbb{F}_p^*}\sum_{z_2\in\mathbb{F}_p^*}
		\zeta_p^{-(z_1\lambda+z_2\lambda_1)}\\&=\begin{cases}
			(p-1)^2p^{m-2},~&\text{if}~\lambda=\lambda_1=0;\\
			-(p-1)p^{m-2},   ~&\text{if}~\lambda=0~\text{and}~\lambda_1\neq 0,~\text{or}~\lambda\neq0~\text{and}~\lambda_1= 0;\\
			p^{m-2}, ~&\text{if} ~\lambda\lambda_1\neq0.
		\end{cases}
	\end{align*}	
	 If $\mathrm{Tr}(ab)\neq0$, it follows from Lemma $\ref{L1.2}$ that
	\begin{align*}
	\Omega_2=&p^{m-2}\sum_{z_1\in\mathbb{F}_p^*}\sum_{z_2\in\mathbb{F}_p^*}
		\zeta_p^{-(z_1\lambda+z_2\lambda_1)}\zeta_p^{-z_1^{-1}z_2^2\mathrm{Tr(ab)}}\\
		=& p^{m-2}\sum_{z_1\in\mathbb{F}_p^*}\zeta_p^{-z_1\lambda}\sum_{z_2\in\mathbb{F}_p^*}
		\zeta_p^{(-z_1^{-1}\mathrm{Tr(ab)}z_2^2-\lambda_1z_2)}\\
		=& p^{m-2}\sum_{z_1\in\mathbb{F}_p^*}\zeta_p^{-\lambda z_1}\bigg(\zeta_p^{(\lambda_1^2(4\mathrm{Tr}(ab))^{-1}z_1)}
		\eta_1(-\mathrm{Tr}(ab)z_1^{-1})G_1-1\bigg)\\
		=& p^{m-2}G_1\sum_{z_1\in\mathbb{F}_p^*}\bigg(\zeta_p^{\big((4\mathrm{Tr}(ab))^{-1}\lambda_1^2-\lambda\big)z_1}
		\eta_1\big(-(4\mathrm{Tr}(ab))^{-1}z_1\big)\bigg)- p^{m-2}\sum_{z_1\in\mathbb{F}_p^*}\zeta_p^{-\lambda z_1} \\
		=& \begin{cases}		
		-(p-1)p^{m-2},&~\text{if}~\lambda=\lambda_1=0;\\	\big(\eta_1(-1)G_1^2-(p-1)\big)p^{m-2},&~\text{if}~\lambda=0~\text{and}~\lambda_1\neq0;\\
			p^{m-2},&~\text{if}~\lambda\neq0~ \text{and}~ 4\lambda\mathrm{Tr}(ab)-\lambda_1^2=0;\\
			\big(\eta_1(4\lambda\mathrm{Tr}(ab)-\lambda_1^2)G_1^2+1\big)p^{m-2},&~\text{if}~\lambda\neq0~ \text{and}~ 4\lambda\mathrm{Tr}(ab)-\lambda_1^2\neq0.
		\end{cases}
	\end{align*}	
	
	So far, we complete the proof for Lemma $\ref{ML}$. $\hfill\Box$\\
	\begin{lemma}\label{ML*}
		For any positive integer $d$, $\lambda_1\in \mathbb{F}_{p}$, $(a,b)\in \mathbb{F}_{p^m}\times\mathbb{F}_{p^m}\backslash\{(0,0)\}$, and
			\begin{align}
		N^{*}_{\lambda_1}(a,b)=\{(x,y)\in \mathbb{F}_{p^m}^*\times \mathbb{F}_{p^m}^*~\big{|}~\mathrm{Tr}(x^{d+1}y)=0~\text{and}~\mathrm{Tr}(ax^dy+bx)=\lambda_1\}.
		\end{align}
     Then, the following assertions hold. 
     
     	$(1)$ For $\lambda_1=0$, 
		\begin{align}\label{5.1*}
		\begin{aligned}
		&\#N^{*}_{\lambda_1}(a,b)\\=&
		\begin{cases}
		p^{2m-2}-2p^{m-1}+1,~&\text{if}~a=0~\text{and}~b\neq0,~\text{or}~\mathrm{Tr}(ab)\neq 0,~\text{or}~a\neq0~\text{and}~b=0;\\		
		p^{2m-2}+(p-3)p^{m-1}+1,~&\text{if}~a\neq0,~b\neq0~\text{and}~\mathrm{Tr}(ab)=0.\\
		\end{cases}
		\end{aligned}
		\end{align}
		
		$(2)$ For $\lambda_1\neq0$, 		
		\begin{align}\label{5.2*}
		\begin{aligned}
		&\#N^{*}_{\lambda_1}(a,b)\\=&
		\begin{cases}
		p^{2m-2}-p^{m-1},~&\text{if}~a=0~\text{and}~b\neq0,~\text{or}~a\neq0~\text{and}~\mathrm{Tr}(ab)\neq 0,~\text{or}~a\neq0~\text{and}~b=0;\\	
		p^{2m-2}-2p^{m-1},~&\text{if}~a\neq0,~b\neq0~\text{and}~\mathrm{Tr}(ab)=0.\\
		\end{cases}
		\end{aligned}
		\end{align}	
     \end{lemma}
 
     {\bf Proof.} By the definitions of both $N_{0,\lambda_1}(a,b)$ and $N^{*}_{\lambda_1}(a,b)$, we have
     \begin{align}\label{Nstar}
     \begin{aligned}
     &\# N^{*}_{\lambda_1}(a,b)\\
     =&\# N_{0,\lambda_1}(a,b)-\# \{(x,y)\in \mathbb{F}_{{p^m}}^*\times\{0\}\big{|}\mathrm{Tr}(bx)=\lambda_1\}\\
     =&\# N_{0,\lambda_1}(a,b)-\begin{cases}
     p^m-1,\quad&\text{if}~b=\lambda_1=0;\\
     p^{m-1}-1,\quad&\text{if}~b\neq0~\text{and}~\lambda_1=0;\\  
     0,\quad&\text{if}~b=0~\text{and}~\lambda_1\neq 0;\\ 
     p^{m-1},\quad&\text{if}~b\neq0~\text{and}~\lambda_1\neq 0.\\ 
     \end{cases}
     \end{aligned}	
     \end{align}			
     Now from Lemma $\ref{ML}$ $(1)$-$(2)$, and $(\ref{Nstar})$, one can get Lemma \ref{ML*} $(1)$-$(2)$, respectively.$\hfill\Box$\\

	Since the trace function is an uniform map, the following lemma is obvious. 
	\begin{lemma}\label{l2}
	    For $t\in \mathbb{F}_p$, denote
	    \begin{align*}
	    	A(t)=\{(a,b)\in\mathbb{F}_{p^m}^*\times\mathbb{F}_{p^m}~\big{|}~\mathrm{Tr}(ab)=t\},
	    \end{align*}
	    then 
	    \begin{align}\label{A}
	    \#	A(t)=p^{m-1}(p^m-1).
	    \end{align}
	\end{lemma}
	
\subsection{The proofs for Theorems $\ref{t1}$-$\ref{t3}$}
{\bf The proof for Theorem $\ref{t1}$}.

 From $(\ref{NDL1})$, $\mathcal{C}_{D_0}$ has length
     \begin{align*}
  	 n=\#D_{0}=p^{2m-1}-p^{m-1}.
     \end{align*} 
     Then, for $\lambda_1\in  \mathbb{F}_p$ and $a,b\in \mathbb{F}_{p^m}$, by Lemma $\ref{ML}$, we have the following two cases.

     {\bf Case 1}. For $a=0$ and $b\neq0$ or $\mathrm{Tr}(ab)\neq0$, 
          \begin{align*}
         \# N_{0,\lambda_1}(a,b)=
          \begin{cases}
          p^{2m-2}-p^{m-1},&\qquad\lambda_1=0;\\
          p^{2m-2},&\qquad\text{otherwise}.
          \end{cases}
          \end{align*}
     Hence each codeword of $\mathcal{C}_{D_0}$ has weight 
      \begin{align*}
      	w= n- \#N_{0,0}(a,b)=(p-1)p^{2m-2},
      \end{align*} and then, by Lemma $\ref{l2}$, the frequency 
         \begin{align*}
         A_w&=(p^m-1)+\sum_{t\in\mathbb{F}_p^*} \#A(t)=(p^m-p^{m-1}+1)(p^m-1).
         \end{align*}
      
      {\bf Case 2}. For $a\neq 0$ and $\mathrm{Tr}(ab)=0$, 
          \begin{align*}
           \#N_{0,\lambda_1}(a,b)=
          \begin{cases}
          p^{2m-2}+(p-2)p^{m-1},&\qquad\lambda_1=0;\\
          p^{2m-2}-p^{m-1},&\qquad\text{otherwise}.
          \end{cases}
          \end{align*}
      Hence each codeword of $\mathcal{C}_{D_0}$ has weight  \begin{align*}
       w=n-\#N_{0,0}(a,b)=(p-1)(p^{m-1}-1)p^{m-1},
       \end{align*} and then by Lemma $\ref{l2}$, the frequency 
       \begin{align*}
       	A_w= \#A(0)=p^{m-1}(p^m-1).
       \end{align*}
      
   So far, we complete the proof for Theorem $\ref{t1}$. $\hfill\Box$\\
      
     {\bf The proof for Theorem $\ref{t2}$}. 
     
     Basing on $(\ref{NDL*})$  and Lemma $\ref{ML*}$, in the similar proof as that for Theorem $\ref{t1}$, one can obtain Theorem \ref{t2} immediately.$\hfill\Box$
 
    {\bf The proof for Theorem $\ref{t3}$}.
    
     For $\lambda\in \mathbb{F}_p^*$, by $(\ref{NDL1})$,  $C_{D_\lambda}$ has length
       \begin{align*}
       n=\#D_{\lambda}=p^{2m-1}-p^{m-1}.
       \end{align*} 
       Then, for $\lambda_1\in  \mathbb{F}_p$ and $a,b\in \mathbb{F}_{p^m}$, by Lemma $\ref{ML}$, we have the following two cases.

       {\bf Case 1}. For $a=0~\text{and}~b\neq0,~\text{or}~a\neq0~\text{and}~\mathrm{Tr(ab)=0}$, it follows from $(\ref{5.3})$-$(\ref{5.4})$ that 
       \begin{align}\label{t31}
        \#N_{\lambda,\lambda_1}(a,b)
       \begin{cases}
       p^{2m-2}-p^{m-1},&\qquad \lambda_1=0;\\
       p^{2m-2},&\qquad\text{otherwise}.
       \end{cases}
       \end{align}
        Hence each codeword of $\mathcal{C}_{D_\lambda}$ has weight 
        \begin{align}\label{w31}
        w=n- \#N_{\lambda,0}(a,b)=(p-1)p^{2m-2},
        \end{align} and then the frequency 
        \begin{align}\label{a31}
        	A_w=(p^m-1)+\#A(0)=(p^{m-1}+1)(p^m-1).
        \end{align}
          
       {\bf Case 2}. For $\mathrm{Tr}(ab)\neq 0$, by $(\ref{5.3})$-$(\ref{5.4})$, we have
              \begin{align}\label{t32}
                \#N_{\lambda,\lambda_1}(a,b)
              = p^{2m-2}+\eta_1(\lambda_1^2-4\lambda\mathrm{T(ab)})p^{m-1}. 	
              \end{align} 
         If $\eta_1(-\lambda\mathrm{Tr}(ab))=-1$, we have
       \begin{align*}
       	  \#N_{\lambda,0}(a,b)=p^{2m-2}-p^{m-1}.	
       \end{align*}
       Thus each codeword of $\mathcal{C}_{D_\lambda}$ has weight 
       \begin{align}\label{w32}
       	w=n- \#N_{\lambda,0}(0,0)=(p-1)p^{2m-2},
       \end{align} and the frequency 
       \begin{align}\label{a32}
       A_w=	\sum_{\eta_1(\lambda t)=1} \#A(t)=\sum_{\eta_1(t)=\eta_1(\lambda)} \#A(t)=\frac{p-1}{2}p^{m-1}(p^m-1).
       \end{align}
	If $\eta_1(-\lambda\mathrm{Tr}(ab))=1$, one has
      \begin{align*}
        \#N_{\lambda,0}(a,b)=p^{2m-2}+p^{m-1}.	
       \end{align*}
        Thus each codeword of $\mathcal{C}_{D_\lambda}$ has weight 
        \begin{align}\label{w33}
        	w=n- \#N_{\lambda,0}(0,0)=(p^{m}-p^{m-1}-2)p^{m-1},
        \end{align} and then the frequency 
        \begin{align}\label{a33}
         A_w=	\sum_{\eta_1(\lambda t)=-1} \#A(t)=\sum_{\eta_1(t)=-\eta_1(\lambda)} \#A(t)=\frac{p-1}{2}p^{m-1}(p^m-1).
       \end{align}
       
      Now, from $(\ref{w31})$-$(\ref{a31})$ and $(\ref{w32})$-$(\ref{a32})$, we can obtain the frequency for the weight  $w=(p-1)p^{2m-2}$ is
       \begin{align}\label{p1}
              A_w=(p^{m-1}+1)(p^m-1)+\frac{p-1}{2}p^{m-1}(p^m-1)=\bigg(\frac{p+1}{2}p^{m-1}+1\bigg)\big(p^m-1\big).
       \end{align} 
       
      So far, by $(\ref{w31})$-$(\ref{p1})$, we complete the proof for Theorem $\ref{t3}$. $\hfill\Box$ 
     \section{Applications}
     
      It is well-known that two-weight linear codes have better applications in secret sharing schemes \cite{JY2006,CC2005},  association schemes \cite{AC1984}, authentication codes \cite{CD2005}, and so on. In particular, projective two-weight codes are very precious as they are closely related to finite projective spaces, strongly regular graphs and combinatorial
      designs \cite{RC1986,CD2018,P1972}. Here, we present the following two applications.
      
      \subsection{Applications for secret sharing schemes}
     The secret sharing schemes is introduced by Blakley \cite{BG} and Shamir \cite{KK} in 1979. Based on linear codes, many secret sharing  schemes are constructed \cite{MJ,JY2006, EJM, MJ93, KD2015}. Especially, for those linear codes with all nonzero codewords minimal, their dual codes can be used to construct secret sharing schemes with nice access structures \cite{KD2015}. 
     
     For a linear code $\mathcal{C}$ with length $n$, the support of a nonzero codeword $\mathbf{c}=(c_1,\ldots,c_n)\in\mathcal{C}$ is denoted by 
     \begin{align*}
     \mathrm{supp}(\mathbf{c})=\{i~|~c_i\neq 0,i=1,\ldots,n\}.
     \end{align*} For $\mathbf{c}_1,\mathbf{c}_2\in\mathcal{C}$, when $\mathrm{supp}(\mathbf{c}_2)\subsetneq\mathrm{supp}(\mathbf{c}_1)$, we say that $\mathbf{c}_1$ covers $\mathbf{c}_2$ . 
     A nonzero codeword $\mathbf{c}\in\mathcal{C}$ is  minimal if it does not cover any other codewords in $\mathcal{C}$. 
     
     The following lemma is very useful for determining minimal codewords.
     \begin{lemma}[Ashikhmin-Barg Lemma \cite{AA}]\label{51}
     	Let $w_{min}$ and $w_{max}$ be the minimal and maximal nonzero weights of the linear code $\mathcal{C}$ over $\mathbb{F}_p$, respectively. If
     	\begin{align*}
     	\frac{w_{min}}{w_{max}}>\frac{p-1}{p},
     	\end{align*}
     \end{lemma}
     then each nonzero codeword of $\mathcal{C}$ is minimal.\\
     
     By Theorems $\ref{t1}$-$\ref{t3}$, if $m\ge 3$, then  
     
     $(1)$ for  $\mathcal{C}_{D_0}$, it holds that
     \begin{align*}
     \begin{aligned}
     \frac{w_{min}}{w_{max}}&=\frac{(p-1)(p^{m-1}-1)p^{m-1}}{(p-1)p^{2m-2}}= 1-\frac{1}{p^{m-1}}	> \frac{p-1}{p};
     \end{aligned}
     \end{align*}
     
     $(2)$ for  $\mathcal{C}_{D^*}$, it holds that
     \begin{align*}
     \begin{aligned}
     \frac{w_{min}}{w_{max}}&=\frac{(p-1)(p^{m-1}-2)p^{m-1}}{(p-1)(p^{m-1}-1)p^{m-1}}= 1-\frac{1}{p^{m-1}}	> \frac{p-1}{p};
     \end{aligned}
     \end{align*}	 
     
     $(3)$ for $\mathcal{C}_{D_\lambda}$, it holds that
     \begin{align*}
     \begin{aligned}
     \frac{w_{min}}{w_{max}}&=\frac{(p-1)p^{2m-2}-2p^{m-1}}{(p-1)p^{2m-2}}= 1-\frac{2}{(p-1)p^{m-1}}	> \frac{p-1}{p}.
     \end{aligned}
     \end{align*}
     
     Hence, by Lemma \ref{51}, all nonzero codewords in $\mathcal{C}_{D_0}$, $\mathcal{C}_{D^*}$ or $\mathcal{C}_{D_\lambda}$ are  minimal. Therefore, their dual codes can be employed to construct secret sharing schemes with interesting access structures.  
     \subsection{Strongly regular graphs with new parameters}
     Some notations and results for strongly regular graphs are given as follows \cite{RC1986}.
     
     A connected graph of $N$ vertices is strongly regular with parameters $(N,K,\lambda,\mu)$, if it is regular with valency
     $K$, and according as the two given vertices are adjacent or non-adjacent, the number of vertices joined to two given vertices is $\lambda$ or $\mu$, respectively.
     
     Let $G=[\mathbf{y}_1,\mathbf{y}_2,\ldots,\mathbf{y}_n]$ be the generator matrix of an $[n,k]$ linear code $\mathcal{C}$ over $\mathbb{F}_p$, where $\mathbf{y}_i\in\mathbb{F}_p^k$ $(i=1,2,\ldots,n)$. Let $\mathbf{V}= \mathbb{F}_p^k$, $\mathbf{O}= \{\langle \mathbf{y}_i\rangle~|~ i = 1,2,\ldots,n\}$ and $\Omega= \{v\in \mathbf{V}~|~ \langle v\rangle\in\mathbf{O}\}$. Define a graph $G(\Omega)$ with vertices based on the vectors in $\mathbf{V}$, and
   any two vertices are joint if and only if their difference is in $\Omega$. By Theorem $3.2$ in \cite{RC1986}, $G(\Omega)$ is strongly regular
     if and only if $\mathcal{C}$ is a projective two-weight code. Assume that the nonzero weights of $\mathcal{C}$ are just $w_1$ and $w_2$. By Corollary $3.7$ in \cite{RC1986}, the parameters of $G(\Omega)$ are given in the following,
     \begin{align*}
     	&N=p^k,\\
     	& K=(p-1)n,\\
     	&\lambda=K^2+3K-p(w_1 +w_2)-Kp(w_1+w_2)+p^2w_1w_2,\\
     	&\mu= K^2+K-Kp(w_1+w_2)+p^2w_1w_2.
     \end{align*}
     
       Now, we give the corresponding strongly regular graphs from $\mathcal{C}_{\overline{D^*}}$ and $\mathcal{C}_{\overline{D_0}}$, respectively.
       
      By Corollaries $\ref{c1}$ and $\ref{c4}$, we know that $\mathcal{C}_{\overline{D_0}}$ is a projective two-weight code with parameters $\big{[}\frac{p^{2m-1}-p^{m-1}}{p-1}  ,2m,(p^{m-1}-1)p^{m-1}\big{]}$ and the weight distribution
      \begin{align*}
      	1+(p^m-1)(p^m-p^{m-1}+1)z^{p^{2m-2}}+(p^m-1)p^{m-1}z^{(p^{m-1}-1)p^{m-1}}.
      \end{align*} 
      Thus, $\mathcal{C}_{\overline{D_0}}$ yields a strongly regular graph $G(\Omega)$ with the following parameters,
      \begin{align*}
      N=p^{2m},~~ K=p^{2m-1}-p^{m-1},~~\lambda=p^{2m-2}+p^m-3p^{m-1},~~\mu=p^{2m-2}-p^{m-1} .
      \end{align*}
      
      Similarly, $\mathcal{C}_{\overline{D^*}}$ can yield a strongly regular graph with the following parameters,
      \begin{align*}\tiny
      &N=p^{2m},~~K=p^{2m-1}-p^m-p^{m-1}-1,\\
      &\lambda=p^{2m-2}+p^m-5p^{m-1}+4,~~\mu=(p^{m-1}-1)(p^{m-1}-2).
      \end{align*}
      
	Compared with the known strongly regular graphs \cite{ZH2021,RC1986,ZH20161,CD2007}, the above strongly regular graphs are new classes.      
   
	 \section{Conclusions}
 
	In this paper, for any odd prime $p$, we construct several classes of two-weight linear codes over $\mathbb{F}_p$ from defining sets, and then obtain two classes of projective two-weight codes $\mathcal{C}_{\overline{D_0}}$ and $\mathcal{C}_{\overline{D^*}}$ by  puncturing $\mathcal{C}_{D_0}$ and $\mathcal{C}_{D^*}$, respectively. All these codes can be suitable for applications in secret sharing schemes with interesting access structures. Furthermore, both $\mathcal{C}_{\overline{D_0}}$ and $\mathcal{C}_{\overline{D^*}}$ can yield new strongly regular graphs, respectively.

       \end{document}